\global\def\draftcontrol{0}

   \def\versionno{ kt inflation -- draft   }

\catcode`\@=11

\expandafter\ifx\csname draftcontrol\endcsname\relax\global\def\draftcontrol{0}
\fi

{\count255=\time\divide\count255 by 60
\xdef\hourmin{\number\count255}
\multiply\count255 by-60\advance\count255 by\time
\xdef\hourmin{\hourmin:\ifnum\count255<10 0\fi\the\count255}}
\def\draftdate{\number\month/\number\day/\number\year\ \ \ \hourmin }

\newcommand\makepapertitle{\par
  \begingroup
    \renewcommand\thefootnote{\@fnsymbol\c@footnote}%
    \def\@makefnmark{\rlap{\@textsuperscript{\normalfont\@thefnmark}}}%
    \long\def\@makefntext##1{\parindent 1em\noindent
            \hb@xt@1.8em{%
                \hss\@textsuperscript{\normalfont\@thefnmark}}##1}%
     \newpage
     \global\@topnum\z@   
     \@makepapertitle
     \thispagestyle{empty}\@thanks
  \endgroup
  \setcounter{footnote}{0}%
  \global\let\thanks\relax
  \global\let\makepapertitle\relax
  \global\let\@makepapertitle\relax
  \global\let\@thanks\@empty
  \global\let\@author\@empty
  \global\let\@date\@empty
  \global\let\@title\@empty
  \global\let\title\relax
  \global\let\author\relax
  \global\let\date\relax
  \global\let\and\relax
  \def\version{\let\version\@version\@gobble}
}
\def\@makepapertitle{%
  \newpage
   \ifnum\draftcontrol=1 {}
   \version\versionno
   \vskip 3em%
   \else
   \hfill\hbox to 3cm {\parbox{4cm}{\@pubnum}\hss}%
   \vskip 3em%
   \fi
   \begin{center}%
   \let \footnote \thanks
     {\LARGE {\@title}}%
     \vskip 1.5em%
     {\normalsize
       \lineskip .5em%
       \begin{tabular}[t]{c}%
         \@author
       \end{tabular}\par}%
     \vskip 1.5em%
     {\@bstract}%
     \end{center}%
     \vskip 1.5em
     \@date%
   \par
}

\gdef\@pubnum{}
\def\pubnum#1{%
  \gdef\@pubnum{#1}}

\gdef\@bstract{}
\def\Abstract#1{%
  \gdef\@bstract{%
   \parbox{\textwidth-0pc}{%
   \centerline{\bf Abstract}\penalty1000%
\kern.2cm%
\noindent
\renewcommand\baselinestretch{1.0}%
{#1}}}
}

\def\ps@paper{\let\@mkboth\@gobbletwo%
     \ifnum\draftcontrol=1
	\def\@oddfoot{\hbox to \textwidth{\tiny \versionno \hfil\tiny\draftdate}%
	\hskip -\textwidth \hbox to \textwidth{\hfil\rm\thepage\hfil}}%
     \else\def\@oddfoot{\hbox to \textwidth{\hfil\rm\thepage\hfil}}
     \fi
     \let\@evenfoot\@oddfoot
}

\def\body{\clearpage
          \pagestyle{paper}
	}

\def\@version#1{\ifnum\draftcontrol=1
\typeout{}\typeout{#1}\typeout{}
\vskip3mm\centerline{\hbox{\fbox{\normalsize{\tt DRAFT -- #1 -- }
                   {\draftdate}}}}\vskip3mm
\fi}
\let\version\@version
\long\def\eqlabel#1{\ifnum\draftcontrol=1
                    \tag@false  
                    \tag*{(\theequation) \hbox to -0.2cm{\hspace{0cm}\small{#1}\hss}}
                    \refstepcounter{equation}
                    \edef\@currentlabel{\theequation}
                    \ltx@label{#1}          
                    \else
                    \label{#1}
                    \fi
                    }
\let\st@bibitem\@bibitem
\let\st@lbibitem\@lbibitem
\ifnum\draftcontrol=1
  \def\@bibitem#1{%
    \st@bibitem{#1}\a@@label{#1}\ignorespaces}
  \def\@lbibitem[#1]#2{%
    \st@lbibitem[#1]{#2}\a@@label{#2}\ignorespaces}
  \def\a@@label#1{%
    \gdef\a@lab{\smash{\normalfont\small#1}}
    \ifvmode
      \if@inlabel
        \global\setbox\@labels\hbox{%
          \llap{\a@lab\let\a@lab\relax
                \kern\@totalleftmargin\kern\marginparsep}%
          \box\@labels}%
      \fi
    \fi}
\fi

\documentclass[12pt,letterpaper]{article}

\usepackage{amsmath,amssymb,array,calc,rotating,epsfig,psfrag}
\usepackage[nosort]{cite}

\ifnum\draftcontrol=1
\fi

\renewcommand\baselinestretch{1.25}
\renewcommand\section{\@startsection {section}{1}{\z@}%
                                   {-3.5ex \@plus -1ex \@minus -.2ex}%
                                   {2.3ex \@plus.2ex}%
                                   {\normalfont\large\bfseries}}
\renewcommand\subsection{\@startsection{subsection}{2}{\z@}%
                                   {-3.25ex\@plus -1ex \@minus -.2ex}%
                                   {1.5ex \@plus .2ex}%
                                   {\normalfont\normalsize\bfseries}}
\renewcommand\subsubsection{\@startsection{subsubsection}{3}{\z@}%
                                   {-3.25ex\@plus -1ex \@minus -.2ex}%
                                   {1.5ex \@plus .2ex}%
                                   {\normalfont\normalsize\it}}
\renewcommand\paragraph{\@startsection{paragraph}{4}{\z@}%
                                   {-3.25ex\@plus -1ex \@minus -.2ex}%
                                   {1.5ex \@plus .2ex}%
                                   {\normalfont\normalsize\bf}}

\numberwithin{equation}{section}

\def\ie{{\it i.e.}}

\def\revise#1       {\raisebox{-0em}{\rule{3pt}{1em}}%
                     \marginpar{\raisebox{.5em}{\vrule width3pt\
                     \vrule width0pt height 0pt depth0.5em
                     \hbox to 0cm{\hspace{0cm}{%
                     \parbox[t]{4em}{\raggedright\footnotesize{#1}}}\hss}}}}

\def\calf         {{\cal F}}

\def\calm         {{\cal M}}

\def\calo         {{\cal O}}

\def\calv         {{\cal V}}

\def\del          {\partial}

\def\sqr#1#2{{\vcenter{\vbox{\hrule height.#2pt
 \hbox{\vrule width.#2pt height#1pt \kern#1pt
 \vrule width.#2pt}\hrule height.#2pt}}}}

\newcommand{\ft}[2]{{\textstyle{\frac{#1}{#2}}}}

\def\om{\Omega}

\def\a{\alpha}

\def\r{\rho}

\def\LL{\Lambda}
\def\ww{\omega}

\def\hg{\hat{g}}

\def\bD3{\overline{D3}}

\def\te{\theta}
\def\p{\phi}
\def\P{\Phi}

\catcode`\@=12

\begin{document}


\title{Inflation in warped geometries}

\pubnum{%
NSF-KITP-03-102 \\
hep-th/0311154}
\date{November 2003}

\author{
Alex Buchel$^{1,2}$ and Radu Roiban$^{3}$\\[0.4cm]
\it $^1$Perimeter Institute for Theoretical Physics\\
\it Waterloo, Ontario N2J 2W9, Canada\\[0.1cm]
\it $^2$Department of Applied Mathematics\\
\it University of Western Ontario\\
\it London, Ontario N6A 5B7, Canada\\[0.1cm]
\it $^3$Department of Physics\\
\it  University of California\\
\it Santa Barbara, CA 93106\\[0.2cm]
}

\Abstract{
We argue that brane anti-brane inflation in 
string theory de-Sitter vacua of Kachru-Kallosh-Linde-Trivedi (KKLT)
is captured by the dynamics of a $D3$-brane probe in the 
local KKLT model constructed in hep-th/0203041. 
This provides a framework to study in a controllable way 
corrections to the inflationary slow roll parameter 
$\eta$ due to  conformal symmetry breaking in a 
warped geometry throat. We compute the leading 
correction to $\eta$ for the inflation in the 
Klebanov-Tseytlin throat geometry. We find 
that in certain regime this correction tends to 
decrease $\eta$.  Computations in a different regime suggest however
that it is unlikely that 
$\eta\ll 1$ can be achieved with the $D3$-brane throat inflation.
}


\makepapertitle

\body

\version\versionno

\section{Introduction}
An important question in  string theory remains
to find an explicit realization  (in a 
single model) of two
main features of our Universe's cosmic evolution:
early inflation and the present-day acceleration.   
Progress in this direction has recently been 
reported in \cite{k2}. Using an effective
four-dimensional description, the  authors  discussed 
embedding of $D3-\bD3$ inflation in the 
string theory de-Sitter vacua of \cite{kklt} (KKLT).
KKLT de-Sitter vacua can be understood by starting with 
string theory flux compactifications of \cite{gkp} (GKP), 
which provides a global description (compactification)
of a local (non-compact) Klebanov-Strassler (KS) model
\cite{ks}. GKP compactifications lead to four dimensional 
 {\it no-scale} models  \cite{ns1,ns2,ns3}: a vanishing 4d
cosmological constant and a complex K\"ahler modulus $\r$  (related to
the overall size of the compactification manifold).   In \cite{kklt} it
was pointed out that string theory instanton corrections \cite{w96}
modify the no-scale structure of GKP, leading to a supersymmetric
$AdS_4$ vacuum with fixed $\r=\r_{c}$. 
It  was further argued that,  for suitable choices of parameters,
the $\bD3$ brane lifts the $AdS_4$ vacuum to a $dS_4$ background, 
while keeping stabilized all moduli of the compactification manifold. 
The $\bD3$ moduli are stabilized as well, as in the 
warped geometry of the compactification manifold it is driven 
to a point where the warp factor has local minimum\footnote{This 
occurs at the end of the Klebanov-Strassler throat of the GKP 
compactification.}. Lastly, the inflation in the KKLT vacua 
is realized by placing a $D3$-brane (a {\it probe}) 
in the throat region of the 
compactification manifold \cite{k2}. In this inflationary 
model the inflaton field $\phi$ of the effective four dimensional 
description is represented by the separation between the probe 
brane and the $\bD3$ brane, stabilized at the end of the throat.
Unfortunately, it was argued that the slow roll parameter 
associated with the $\phi$-field inflation is too large
for this model to be realistic
\begin{equation}
\eta\equiv \frac 13 \frac{V_{\it inf}(\phi)''}{H^2}=\frac 23\,.
\eqlabel{e0}
\end{equation}
In \eqref{e0} $V_{\it inf}(\phi)$ is the inflaton potential obtained
after  
integrating out the K\"ahler modulus of the compactification 
manifold
\begin{equation}
V_{\it inf}=V(\r,\phi)\bigg|_{\r=\r_c}\,,
\eqlabel{vinf}
\end{equation}
 and $H$ is the Hubble scale of the de-Sitter vacua.  It was suggested 
\cite{k2} that the generic $\eta$-problem \eqref{e0} might be alleviated 
once the $\phi$-dependence of the $\r$-modulus superpotential is taken
into account, or if a K\"ahler stabilization mechanism (as opposite to
the 
superpotential stabilization) is used to fix $\r$. However, 
it is not known whether either of these mechanisms actually works:
the $\phi$-dependence of the effective four dimensional 
superpotential is not known at present
and, though some of the $\a'$ corrections that break the no-scale 
scale structure of GKP are computed \cite{ac}, the reported 
corrections do not lead to the suggested K\"ahler stabilization 
mechanism. 

In this paper we explain a different framework for analyzing 
inflation in warped de-Sitter string theory geometries 
which, in particular, bypasses the difficulties of computing corrections 
to $\eta$ from the effective four dimensional perspective 
mentioned above. We observe that for some aspects of the brane inflation 
deep inside the warped throat geometries the details of the
compactification manifold, which provides a UV completion of the 
otherwise infinite throat, are irrelevant. All that matters 
from the compactification manifold is that it supplies a 
four dimensional Hubble parameter $H$. Also, in this 
setup it is assumed that all moduli of the compactification 
manifold are fixed, and the scale of moduli stabilization
$E_s$ is much higher then the relevant scales  of inflation
$E_s\gg H$, $E_s\gg |\phi|$. It is clear that 
brane inflation in this class of models is equivalent to the 
$D3$ probe brane dynamics in the local geometry where the throat, 
rather then terminating on some complicated (compact) Calabi-Yau
manifold, extends to infinity. The advantage of this viewpoint 
is that, unlike compact KKLT or GKP backgrounds, the corresponding 
local models can be rather easily and explicitly 
constructed. For example, much like 
KS model \cite{ks} is a local description of the throat geometry 
of the GKP compactification, the de-Sitter deformed  KT model
\cite{kt} 
described in \cite{bt,b1} is a local realization of the 
throat geometry of the KKLT model\footnote{
Strictly speaking, the correct local model would be de-Sitter 
deformation of the Klebanov-Strassler solution \cite{ks}.
For the inflation occurring far from the 
end of the KS throat the difference between 
KT and KS models is subdominant, as it will be for their 
corresponding de-Sitter deformations. KS de-Sitter 
deformation as proposed in \cite{b1} can be explicitly constructed
\cite{bu}. 
}. The inflation, or equivalently 
the brane probe dynamics, can now be studied very explicitly
and analytically. 

In the next section we review the effective four
dimensional approach to inflation in KKLT de-Sitter 
vacua \cite{k2}. In section 3, after explaining the 
general relation between effective four dimensional inflation 
and higher dimensional probe brane dynamics, 
we show that the $\eta$-problem 
\eqref{e0} arises in the leading approximation 
from the $D3$ probe dynamics in $AdS_5\times T^{1,1}$ 
(KW model \cite{kw}), with $AdS_5$ written in 
hyperbolic coordinates (or de-Sitter slicing). 
We then study $D3$ probes in the de-Sitter deformed 
KT geometry  and demonstrate that in certain regime 
$\eta<\frac 23$. We point out  that it is unlikely that
$\eta\ll 1$ can be achieved in this class of inflationary models.
 
We would like to emphasize that though in this paper we study 
inflation as a probe dynamics in  de-Sitter deformed
KT backgrounds \cite{bt,b1}, our approach is quite general. 
Following the prescription given in \cite{b1}, de-Sitter deformation 
can be constructed in principle for {\it any} RG flow 
realizing supergravity dual to a four dimensional gauge theory. 
Some deformations of this type are discussed in \cite{blw,b2}.
It is possible to  study analytically the probe dynamics in these 
backgrounds, and try to address the question whether 
always $\eta<\frac 23$, and in particular, whether it is possible 
to achieve $\eta\ll 1$. Also, one can study more exotic inflationary
models,  as a probe dynamics of a $Dp$-brane, for $p>3$, 
wrapping a $(p-3)$ cycle of the transverse manifold. An example 
of this would be a five-brane wrapping a two-cycle of the 
de-Sitter model described in section 3 of \cite{blw}.
 
\section{4d effective description of inflation}

Consider a set of parallel branes placed at different positions in
their common transverse space. Their relative position is described by
some matrix-valued scalar field living on their world volume. Under
suitable circumstances this theory also contains dynamical gravity.
The idea of $D-{\overline D}$ inflation \cite{td} is that in such a system of
branes and anti-branes, the field describing their separation  
can act as an inflaton in the world volume theory. Because the system
is not BPS, the branes experience a net attractive force and thus
the would-be inflaton has a nontrivial potential. For branes in flat space
this potential leads to a large slow roll parameter, which makes this
scenario somewhat unsuitable for sustained slow roll inflation.

This problem is slightly alleviated by embedding \cite{k2}
the mechanism described above in a warped compactification, 
the intuition being that the warping of the geometry leads to an
inflaton potential which is flatter 
than in the case of branes in flat space. This intuition is realized
if one ignores the stabilization of the modulus corresponding to the
size of the transverse space. Taking it into account modifies this
conclusion and leads to fast roll inflation \cite{k2}. We will now
briefly review these features.

The KKLT scenario is based on a compact geometry.
To leading order, the local geometry (Klebanov-Strassler throat)
can be approximated by  $AdS_5\times X^5$, with the usual
assumptions leading to small curvature:
\begin{equation}
ds^2=h(r)^{-1/2}(-dt^2+d{\vec x}^2)+h(r)^{1/2}(dr^2 + r^2
ds^2_{X^5})\,,~~~~~~~h(r)=\frac{R^4}{r^4}\,. 
\end{equation} 
At the IR and UV ends of the geometry there are significant departures
from this structure. In particular, in the IR the geometry is smoothly
cutoff before $h$ blows up; this leads to the non-decoupling of gravity
and the existence of a standard Einstein-Hilbert action in the
four-dimensional effective action.

Due the the cancellation between the gravitational attraction and
the electrostatic repulsion due to the RR flux, a $D3$ brane
experiences no force. A $\bD3$ brane however
is dynamically localized at the IR end of the AdS throat. Combining these
two effects leads to a relatively shallow potential felt by the brane,
which is entirely due to the presence of the anti-brane and is
substantially different from the flat space one due to the warping of
the geometry.  This potential, which in the compact model appears in
the four-dimensional effective action,  is computed
in a probe approximation; denoting by $r_0$ and $r_1$ the location of the
anti-brane and the distance to the brane, respectively, the potential
turns out to be 
\cite{k2}  
\begin{equation}
V=2T_3
\frac{r_0^4}{R^4}\left(1-\frac{1}{N}\frac{r_0^4}{r^4_1}\right)\,.
\label{Vs}
\end{equation}
In explicit string theory realizations of this scenario the
coefficient $r_0/R$ is exponentially small and thus (\ref{Vs}) leads 
to the required flat inflaton potential.

In the presence of a stabilized radial modulus the situation is
somewhat different and \cite{k2} argues that the $\eta$-parameter
 is always of order unity. At the level of the effective action
this is due to an unfortunate interplay between the would-be inflaton 
$\phi$ (the analog of $r_1$ in equation (\ref{Vs})) describing the
position of the $D3$ brane in the KS throat, the 
four-dimensional field $\rho$ associated to the K\"ahler form and the
correct identification of the actual volume modulus $r$. As argued in 
\cite{dg}, the four-dimensional K\"ahler potential of the combined
$(\phi,\,\rho)$ system is 
\begin{equation}
K(\rho, \phi)=-3
\ln(\rho+{\bar \rho}-k(\phi,{\bar \phi}))\,,
\end{equation}
which leads to the identification of the actual volume modulus $r$ as
\begin{equation}
2r=\rho+{\bar \rho}-k(\phi,{\bar \phi})\,.
\end{equation}
The complete potential now has two possible rolling directions.
For artificially fixed volume modulus the surviving direction
exhibits slow roll. If however,  the volume modulus is dynamically
fixed using a superpotential-based stabilization mechanism, 
the surviving direction exhibits fast roll\footnote{The problem
reviewed here can in principle be 
bypassed \cite{k2} by using a K\"ahler-based stabilization
mechanism, but such compactifications are difficult to control.}.
This is mainly due to the fact that this mechanism yields an
expectation value for $\rho$ rather than $r$. The problem emerges from
the fact that, as  $\rho$ acquires a nonvanishing expectation value,
the would-be canonically normalized inflaton
\begin{equation}
\varphi=\phi\sqrt{6/(\rho+{\bar\rho})}
\label{can_infl}
\end{equation}
acquires a mass of the order
of the vacuum energy which in turn leads to an $\eta$-parameter of
order unity. Since in the following section we will describe
an alternative to the discussion in \cite{k2}, let us describe this in
more detail.

Given an arbitrary superpotential which depends only on $\rho$, 
the potential for the combined $(\phi,\,\rho)$ system is
\begin{equation}
V=\frac{1}{6r}\left[\partial_\rho W {\bar\partial_\rho W}\left(1+
\frac{1}{2r}\frac{\partial_\phi k \partial_{\bar \phi}k}
{\partial_{\phi {\bar \phi}}k}\right)-\frac{3}{2r} \left({\bar
W}\partial_\rho W + W {\bar \partial_\rho W}\right) \right]+V_{\it
anti-brane}\,.
\label{pot}
\end{equation}
In this equation, the square bracket arises as a consequence of
the existence of the superpotential $W$ while the last term is due to
the presence of the anti-brane at the end of the throat. It is worth
pointing out that this last term is analogous to the first term in
(\ref{Vs}) while the second term in that equation is part of the
square bracket in the equation above.

Assuming that $k=\p\overline\p$, the potential (\ref{pot}) has a de-Sitter minimum at
some $\rho=\rho_c$ and vanishing $\phi$ at which the potential takes
the value $V_c=V(\rho_c, 0)$ and defining the canonically-normalized
field (\ref{can_infl}), it follows that \cite{k2}
\begin{equation}
V\simeq V_c\left(1+\frac{1}{3}\varphi{\bar \varphi}\right)\,.
\end{equation}
This implies that the ``inflaton'' field $\varphi$ behaves similarly
to a conformally coupled scalar in a space with cosmological constant 
$V_c$
\begin{equation}
m^2_{\varphi}=\frac{1}{3}V_c=H^2\,,
\label{mass}
\end{equation}
where $H$ is the Hubble constant. This implies that $\eta$ is of order
unity and thus incompatible with sustained slow roll inflation.

In this effective field theory framework it is difficult to analyze
how the inflaton potential responds to deviations  in the geometry
from the exact $AdS$ throat.
In the next section  we will describe an alternative derivation of 
(\ref{mass}) which will allow us to compute such corrections.

\section{Inflation as a probe dynamics}

Recall that after integrating out all moduli of the 
compactification manifold $\calm_6$, the low energy effective 
action is given by \cite{kklt}
\begin{equation}
S_{bulk}=\frac{1}{2k_4^2}\int_{\calm_4}d^4x\sqrt{-g}\biggl(
R-2\Lambda\biggr)\,,
\eqlabel{4d}
\end{equation}    
where $\LL\equiv 3H^2>0$ is the four dimensional 
cosmological constant.
The $D3-\bD3$ inflation of \cite{k2} can be understood 
as an effective description of a $D3$ brane probe dynamics 
in $\calm_{10}=\calm_4\times \calm_6$ warped string theory flux background. 
To identify the inflaton field let us focus 
on the throat geometry of $\calm_6$. 
Locally, the ten dimensional metric is 
\begin{equation}
\begin{split}
ds_{\calm_{10}}^2=&\om_1^2\left(r,\frac{y_i}{r}\right)ds_{\calm_4}^2(x)
+\om_2^2\left(r,\frac{y_i}{r}\right)
ds_{\calm_6}^2(y)\\
=&\om_1^2\left(r,\frac{y_i}{r}\right)\left(dS_4\right)^2+
\om_2^2\left(r,\frac{y_i}{r}\right) dr^2+ ds_5^2\left(r,\frac{y_i}{r}
\right)+\sum_{i=1}^6 g_{ri}\left(r,\frac{y_i}{r}
\right)\ dr dy^i\,,
\end{split}
\eqlabel{10d}
\end{equation} 
where $r$ is the ``radial'' coordinate along the throat, 
roughly $r^2\sim\sum y_i^2$, $ds_5^2$ is the 
``angular'' part of the metric of $\calm_6$,  
and the four dimensional de-Sitter 
space $\calm_4$ has Ricci curvature $12 H^2$. In the ``throat
approximation'' 
\begin{equation}
\om_i\left(r,\frac{y_i}{r}\right)\approx \hat\om_i(r)\,,\qquad 
g_{ri}\approx 0\,.
\eqlabel{throat}
\end{equation}
The metric ansatz approximation \eqref{throat}
implies that the four-form potential\footnote{We 
assume that there is a self-dual five form flux $F_5=(1+\star)\calf_5$,
$\calf_5=dC^{(4)}$, supporting the background geometry 
\eqref{10d}.} $C^{(4)}$ also depends on 
$r$ only
\begin{equation}
C^{(4)}\approx \ww(r)\ vol_{\calm_4}\,,
\eqlabel{cr}
\end{equation} 
where $vol_{\calm_4}$ is a volume form on $\calm_4$.
A $D3$ brane at $r=r_1$ in the throat geometry \eqref{throat}, \eqref{cr}
has an effective 10d description 
\begin{equation}
S_{D3}=-T_3\int_{\calm_4}d^4\xi\sqrt{-\hg(r_1)}+T_3\int_{\calm_4}C^{(4)}
(r_1)\,,
\eqlabel{d3}
\end{equation}  
where $\hg$ is the induced metric on the world-volume of the probe.
For slowly varying $r_1=r_1(x)$, we find  
\begin{equation}
\sqrt{-\hg}=\sqrt{-g}\ \hat\om_1^4(r_1)\ \left(1+\frac 12\
\hat\om_1^{-2}(r_1)\hat\om_2^{2}(r_1)\ 
g^{\mu\nu}
\del_\mu r_1\del_\nu r_1\right)\,,
\eqlabel{slow}
\end{equation}
where $g_{\mu\nu}=g_{\mu\nu}(x)$ is the bulk metric of \eqref{4d}.
Given \eqref{slow}, the probe action evaluates to 
\begin{equation}
S_{D3}=\int_{\calm_4}d^4x\sqrt{-g}\left(-\frac 12 T_3\hat\om_1^2(r_1)
\hat\om_2^2(r_1)\ (\del r_1)^2-\calv(r_1)\right)\,,
\eqlabel{d31}
\end{equation} 
where $\calv(r_1)$ is the $D3$ probe potential
\begin{equation}
\calv(r_1)=T_3\biggl(\hat\om_1^4(r_1)-\ww(r_1)\biggr)\,.
\eqlabel{v}
\end{equation}  
In a consistent Kaluza-Klein reductions, the four dimensional 
effective description of a probe is given by the {\it same} effective 
action\footnote{This has been implicitly assumed for the $\bD3$ 
four-dimensional effective action in the construction of KKLT de-Sitter vacua.
Details that such  a prescription is indeed correct 
will appear in \cite{b3}.} as \eqref{d3}, 
or in the slow roll approximation \eqref{d31}.  
Thus identifying the effective inflaton field $\phi$
with the radial coordinate $r_1$ of the probe brane in the throat
geometry, or more precisely 
\begin{equation}
d\phi\equiv\sqrt{T_3}\hat\om_1(r_1)\hat\om_2(r_1)\ dr_1\,,
\qquad \calv_{\it inf}(\phi)=\calv(r_1(\phi))\,,
\eqlabel{inf}
\end{equation} 
we obtain the following effective inflaton action
\begin{equation}
S_{\it inf}=\int_{\calm_4}d^4x\sqrt{-g}\left(-\frac 12 (\del\phi)^2-
\calv_{\it inf}(\phi)\right)\,.
\eqlabel{sinf}
\end{equation}
Altogether, the complete action $S_{\it cosm}$ describing the
cosmology of \cite{k2}  
is obtained from \eqref{4d}, \eqref{sinf}
\begin{equation}
S_{\it cosm}=S_{\it bulk}+S_{\it inf}\,.
\eqlabel{cosm}
\end{equation}
Notice that, because of gauge invariance, $\ww(r_1)$ (and thus 
$V_{\it inf}(\phi)$) are defined up to an arbitrary additive constant. 
In what follows we fix this constant in such a way 
that 
\begin{equation}
\calv_{\it inf}(\phi)\bigg|_{\phi=0}=0\,.
\eqlabel{const}
\end{equation}
Eq.\eqref{const} implies that the scale of inflation is given 
by the bulk Hubble parameter $H$. In particular, the usual 
slow-roll parameters $\epsilon, \eta$ are given by 
\begin{equation}
\begin{split}
\epsilon=&\frac{k_4^2}{18}\ \left(\frac{V_{inf}'}{H^2}\right)^2\,,\\
\eta=&\frac13\ \frac{V_{inf}''}{H^2}\,,
\end{split}
\eqlabel{slowr}
\end{equation}
where the derivatives are with respect to $\phi$. An important 
observation is that, even  though $\epsilon$
in \eqref{slowr} depends on the four dimensional gravitational 
coupling\footnote{This implies, in particular,  
that $\epsilon$ depends on the details of the compactification
(\ie, volume) 
to the extend that scales $k_4^{-1}$ and the Hubble 
scale $H$ can be correlated.} $k_4$, 
the slow-roll parameter $\eta$ is independent of it. 
Thus,  one can use the exact local (noncompact) description of the 
throat geometries to deduce model-independent predictions 
for $\eta$. For example, up to the comment in footnote 2, 
the exact local throat geometry for the class of inflationary 
models considered in  \cite{k2} was presented in \cite{bt,b1}.

The rest of this section is organized as follows. 
We first demonstrate that the leading contribution $\eta=\frac 23$
follows from the dynamics of a $D3$ probe brane in the exact
$AdS_5\times T^{1,1}$ throat, where $AdS_5$ space 
has four-dimensional de-Sitter 
slicing of curvature $12 H^2$. We then proceed to study probe dynamics in 
the de-Sitter deformed KT solutions presented in \cite{bt,b1}.     
As advocated in \cite{b1}, de-Sitter deformations of the 
supergravity RG flows can be interpreted as holographic dual to 
four dimensional gauge theories in $dS_4$ background space-time.
The fact that $\eta=\frac 23$ for the de-Sitter deformed KW solution \cite{kw} 
simply follows from the fact that the inflaton field $\phi$ is
holographically dual to the conformally coupled  scalar field $\Phi$ 
of the $SU(N)\times SU(N)$ KW gauge theory, which necessarily has 
a coupling\footnote{Analogous observation was also made in \cite{k2}.}
\begin{equation}
V_{\Phi}=\frac {1}{12}\ R_4\ \Phi^2=H^2\ \Phi^2\,,
\eqlabel{confcoupl}
\end{equation}  
where $R_4=12 H^2$ is the Ricci scalar of the background 
de-Sitter space-time of the deformed KW gauge theory. 
The KT/KS gauge theory \cite{kt,ks} can be understood as a deformation of the 
KW gauge theory $SU(N)\times SU(N)\to SU(N+M)\times SU(N)$, 
which induces the effective running of $N$ with scale \cite{kt,bh}  
\begin{equation}
N=N(\mu)=N_0+\frac{3}{2\pi} g_s M^2\ln\frac{\mu}{\mu_0}\equiv 
N_0\biggl(1+P^2\ln\frac{\mu}{\mu_0}\biggr)\,,
\eqlabel{nrunning}
\end{equation}
where $\mu_0$ is the strong coupling scale of the cascading 
gauge theory, $g_s$ is the string coupling\footnote{We set 
$g_s^{-1}\equiv 4\pi(g_1^{-2}(\mu)+g_2^{-2}(\mu))\to 1$ as $\mu\gg
\mu_0$. In the latter 
equation $g_i$ are gauge couplings of the KT gauge groups.},
and $P^2\equiv 3 M^2/(2\pi N_0)$.   
We will study two different regimes in the de-Sitter deformed 
KT gauge theory\footnote{The general case can be studied as well, 
albeit numerically.}
\begin{equation}
\begin{split}
&(a):\quad H\gg \mu_0\ \quad \Longleftrightarrow\qquad 
0<P^2\ln\frac{\mu}{\mu_0}\ll 1\,,\\
&(b):\quad H\ll \mu_0\ \quad \Longleftrightarrow\qquad 
P^2\ln\frac{\mu}{\mu_0}\gg 1\,.
\end{split}
\eqlabel{cases}
\end{equation} 
The background geometry of the supergravity dual 
in the first case, $(a)$, is the leading $P^2$ 
deformation of the KW model discussed in \cite{bt}, 
while the latter case is the leading $H^2$ deformation 
of the KT background geometry which we discuss below. 
In each case we find correspondingly 
\begin{equation}
\begin{split}
&(a):\quad \eta=\frac 23 \left(1-\frac {P^2}{12}\ln\frac{\phi^2}{ L^4
H^2 T_3}
\right),\quad 
0<\frac {P^2}{12}\ln\frac{\phi^2}{L^4 H^2 T_3}\ll 1,\quad
\frac{L^4H^2T_3}{\p^2}\ll 1\,,
\\
&(b):\quad \eta=\frac 23
\left(1+\calo\left(\frac{L^8 H^4 T_3^2}{\p^4}\right)\right),\quad 
\frac {P^2}{12}\ln\frac{\phi^2}{L^4 H^2 T_3}\gg 1,\quad 
\frac{L^4H^2T_3}{\p^2}\ll 1\,,
\end{split}
\eqlabel{ceta}
\end{equation}
where $L=4\pi N_0(\a')\ft {27}{16} $.
The computation of $\eta$ in case $(b)$ suggests that it is unlikely 
that one can achieve (in a computationally controllable way) 
$\eta\ll 1$ for inflationary models in throat geometries of the
KKLT de-Sitter vacua. The point is simply that to have computational 
control one has to study $D3$ probes far from the IR end of the 
Klebanov-Strassler throat, or to be in the regime $(b)$. 
There is a physical reason for this as well. Following 
KKLT construction, there is a $\bD3$ brane at the end of the 
KS throat, and thus bringing a $D3$ probe close to it
will cause strong $D3-\bD3$ attraction and the effective 
four-dimensional inflaton will start rolling too fast.
  
\subsection{Inflation in de-Sitter deformed KW geometry}
Following \cite{b1}, the throat geometry in this case is 
given by 
\begin{equation}
ds_{10}^2=\frac{r^2}{L^2}\left(-dt^2+e^{2H t}d\bar{x}^2\right)
+\frac{ L^2 dr^2}{  L^4 H^2 +r^2}+ L^2 ds_{T^{1,1}}^2\,,
\eqlabel{kw}
\end{equation}
where $(ds_{T^{1,1}})^2$ is the standard metric on 
$T^{1,1}=(SU(2)\times SU(2))/U(1)$ and
\begin{equation}
L^4=4\pi g_s N(\a') \frac{27}{16}\,,
\eqlabel{ldef}
\end{equation}
with $N$ being the number of  $D3$ branes.
The metric \eqref{kw} is supported by the following 5-form
flux
\begin{equation}
F_5={\cal F}_5+\star{\cal F}_5\,,\qquad {\cal F}_5={-4 L^4} 
d{\rm vol}_{T^{1,1}}\,.
\eqlabel{frkw}
\end{equation}
From \eqref{frkw} we conclude that 
\begin{equation}
dC^{(4)}=\frac{4r^4}{L^4\sqrt{L^4 H^2+r^2}}\ dr\wedge dvol_{dS_4}\,.
\eqlabel{c4kw}
\end{equation}
Using \eqref{v}, \eqref{kw}, \eqref{c4kw} we find 
\begin{equation}
\begin{split}
T_3^{-1}\calv(r_1)=&\frac{r_1^4}{L^4}\left(1-\sqrt{1+\frac{L^4H^2}
{r_1^2}}\right)
+\frac 32 H^2 r_1^2 \sqrt{1+\frac{L^4H^2}{r_1^2}}\\
&-\frac 32 L^4 H^4 
\ln\frac{r_1+\sqrt{L^4H^2+r_1^2}}{L^2H}\\
=&H^2 r_1^2\left(1+\calo\left(\frac{L^4H^2}{r_1^2}\right)\right)\,.
\end{split}
\eqlabel{vkt}
\end{equation}
Defining an effective inflaton field as in \eqref{inf} 
\begin{equation}
d\phi=\sqrt{T_3}\frac{r_1}{\sqrt{L^4H^2+r_1^2}}\ dr_1\,,
\eqlabel{fkw}
\end{equation}
we find
\begin{equation}
\begin{split}
\calv_{inf}(\phi)&=H^2\phi^2\left(1+\calo\left(\frac{L^4H^2T_3}{\phi^2}
\right)\right)\\
&\approx H^2\phi^2 \,,
\end{split}
\eqlabel{vinfkw}
\end{equation}
where in the second line we imposed constraint for inflation to 
occur far from the IR end of the throat
\begin{equation}
\frac{L^4H^2T_3}{\phi^2}\ll 1\,.
\eqlabel{irend}
\end{equation}
With \eqref{slowr}, \eqref{vinfkw} we reproduce 
\begin{equation}
\eta=\frac 23\,.
\eqlabel{etakw}
\end{equation}

\subsection{Inflation in de-Sitter deformed KT geometry} 
Without loss of generality we set $L=1$. From \cite{bt,b1}
we extract the exact background describing the throat 
geometry
\begin{equation}
ds^2_{10} =  e^{2z} (dM_4^H)^2
+ e^{-2z}  [e^{10y} du^2 + e^{2y} (dM_5)^2] \,.    
\eqlabel{ktmet}
\end{equation}
Here   $M_5$ is a deformation of the $T^{1,1}$ metric
\begin{equation}
\begin{split}
&(dM_5)^2 =  e^{ -8w}  e_{\psi}^2 +  e^{ 2w}
\left(e_{\theta_1}^2+e_{\phi_1}^2 +
e_{\theta_2}^2+e_{\phi_2}^2\right)  \ , \\
&e_{\psi} =  \frac 13 (d\psi +  \cos \theta_1 d\phi_1
 +  \cos \theta_2 d\phi_2)  \  ,
 \quad  e_{\theta_i}=\frac{1}{\sqrt 6} d\theta_i\ ,  \quad
  e_{\phi_i}=
\frac{1}{ \sqrt 6} \sin\theta_id\phi_i \ ,
\end{split}
\eqlabel{mdet}
\end{equation}
and 
\begin{equation}
(dM_4^H)^2=-dt^2+e^{2 H t} d\bar{x}^2\,.
\eqlabel{m4h}
\end{equation}
As for the matter fields, we have a dilaton $\Phi$ that 
depends on the radial coordinate $u$ only, and the $3-$ and $5-$form 
fluxes
\begin{equation}
\begin{split}
&F_3 = \   P
e_\psi \wedge
( e_{\theta_1} \wedge e_{\phi_1} -
e_{\theta_2} \wedge e_{\phi_2})\ ,
\ \ \ \ \ \ \
B_2  = \    f(u)
( e_{\theta_1} \wedge e_{\phi_1} -
e_{\theta_2} \wedge e_{\phi_2})
 \ ,\\
&F_5= {\cal F}+*
{\cal F}\
, \quad  \ \ \ \ {\cal F} =
K(u)
e_{\psi}\wedge e_{\te_1} \wedge
e_{\p_1} \wedge
e_{\te_2}\wedge e_{\p_2}\  , \ \ \ \ \ \ \
 K (u)  =4+ 2 P f (u) \ .
\end{split}
\eqlabel{fluxes}
\end{equation}
The corresponding system
of type IIB supergravity equations of motion describing the
radial evolution of the five unknown functions of
$u$: $y,z,w,K,\Phi$ is given by \cite{b1}
\begin{equation}
\begin{split}
&10y'' - 8 e^{8y} (6 e^{-2w} - e^{-12 w})  -
 30  H^2  e^{10y-4z }
 + \P''
=0 \ ,\\
&10w'' - 12 e^{8y} ( e^{-2w} - e^{-12 w})   - \P''
=0 \ , \\
&\P''    + e^{-\P + 4z - 4y-4w} (\frac{K'^2}{ 4 P^2} -
 e^{2 \P + 8 y+8w} P^2)=0 \ ,\\
&4z'' -  K^2  e^{8z}
 - e^{-\P + 4z - 4y-4w} ( \frac{K'^2}{ 4 P^2} +
 e^{2 \P + 8 y+8w} P^2)
  - 12  H^2  e^{10y-4z }   =0\ ,\\
&(e^{-\P + 4z - 4y-4w} K')' - 2P^2 K e^{8z} =0 \ ,
\end{split}
\eqlabel{kteqs}
\end{equation}
with the first order  constraint
\begin{equation}
\begin{split}
&5  y'^2    - 2 z'^2  - 5 w'^2 - \frac 18 \P'^2
- \frac 14  e^{-\P +  4z -4y - 4 w }\frac{K'^2}{4 P^2} \\
&-  \ { 3 } H^2  e^{10y -4z}
-    e^{8y} ( 6 e^{-2w} - e^{-12 w} )
 +  \frac 14 e^{\P+  4z + 4y + 4 w } P^2 +
 \frac 18  e^{8z} K^2   = 0  \ .
\end{split}
\eqlabel{constr}
\end{equation}
In eqs.~\eqref{kteqs}, \eqref{constr} prime denotes  derivative with
respect to $u$.

Rather than computing  the $D3$ probe brane potential 
\eqref{v} in the above geometry, it is convenient to compute its 
derivative with respect to the radial coordinate 
$u$. We find
\begin{equation}
\frac{d\calv(u)}{du}=T_3 e^{4z}\biggl(4z'+K e^{4z}\biggr)\,.
\eqlabel{derpot}
\end{equation}
Also, using \eqref{inf}, the effective inflaton field $\p$ is defined
as 
\begin{equation}
d\p=-\sqrt{T_3}\ e^{5y}\ du\,.
\eqlabel{infkt}
\end{equation}

\subsubsection{Case $(a)$}
In order to make use of the results of \cite{bt} we set $H=1$.
The $H$ dependence in the final expression for $\eta$ can 
be restored from dimensional analysis. It will also be convenient 
to introduce a new radial coordinate $\r$ such that
\begin{equation}
d\r=-e^{4y}du\,.
\eqlabel{ru}
\end{equation}
The  $\calo(P^2)$ solution to \eqref{kteqs} can be parameterized 
as 
\begin{equation}
\begin{split}
&K = 4 + 2 P^2 k(\r)
 \ , \ \ \ \ \ \P  = P^2 \hat\P(\r) \ , \ \  \ \ 
 w = P^2 \omega (\r) \ , \\
&y= y_0(\r)   +  P^2 \xi (\r) \ ,\ \ \ \ \ \ \
 \ \ \   z=  y_0(\r)   +  P^2
\eta (\r) \ , \ \ \ \   \  \ \ \ \ \ 
  y_0(\r) \equiv \ln \sinh \r\,.
\end{split}
\eqlabel{lkt}
\end{equation}
Thus, using \eqref{ru} we find from \eqref{derpot}
\begin{equation}
\begin{split}
T_3^{-1} \frac{d\calv(\r)}{d\r}=&4 e^{4y_0}\biggl\{
y_0'\left(1+4P^2\eta\right)+P^2\eta'-1-P^2\left(
\frac k2+8\eta-4\xi\right)+\calo(P^4)
\biggr\}\\
=&4 e^{4y_0}\biggl\{
P^2\left(
\eta'+4\xi-4\eta-\frac k2
\right)+\frac 12 e^{-2y_0}(1+4P^2\eta)+\calo(e^{-4y_0})
\biggr\}+\calo(P^4)\,,
\end{split}
\eqlabel{d1}
\end{equation}
where the derivatives are taken with respect to $\r$ now.
In the second line in \eqref{d1} we used
\begin{equation}
y_0'=\sqrt{1+e^{-2y_0}}=1+\frac 12 e^{-2y_0}+\calo(e^{-4y_0})\,.
\eqlabel{h1}
\end{equation}
Introducing $\nu\equiv5\xi-2\eta$ we find
\begin{equation}
\begin{split}
\eta'+4\xi-4\eta-\frac k2=&-\frac 12 \left(\nu'-4\nu+k\right)+\frac
52\xi'
-6\xi\\
=&\frac 13 e^{-2y_0}\ \nu+\cdots\\
=&\frac {1}{12}e^{-2y_0}\r+\cdots\,,
\end{split}
\eqlabel{h2}
\end{equation}
where we used eqs.~(4.30), (4.31), (4.36) of ref.~\cite{bt}, 
and the asymptotics  (eq.~(4.18) of ref.~\cite{bt})
\begin{equation}
\begin{split}\
&\nu\to \frac 14\r\,,\qquad \r\to\infty\,,\\
&\eta\to -\frac 18\r\,,\qquad \r\to\infty\,,\\
&\xi\to -\frac 16 e^{-2y_0}\nu\,,\qquad \r\to\infty\,.  
\end{split}
\eqlabel{h3}
\end{equation}
The ellipses in \eqref{h2} 
denote subdominant terms as $\r\to \infty$.
With \eqref{h2}, \eqref{h3} we conclude from \eqref{d1}
\begin{equation}
T_3^{-1} \frac{d\calv(\r)}{d\r}=2 e^{2 y_0}\left(
1-P^2\left(\frac{\r}{6}+\calo(1)\right)+\calo(e^{-2y_0})
\right)+\calo(P^4)\,.
\eqlabel{d2}
\end{equation}
Using \eqref{ru}, the inflaton is defined according to \eqref{infkt}
\begin{equation}
\begin{split}
T_3^{-1/2} d\p=&e^{y}\ d\r\\
=&e^{y_0}(1+P^2\xi+\calo(P^4))\ d\r\\
=&e^{y_0}(1-\frac{P^2}{24}e^{-2y_0}\r+\calo(P^4))\ d\r\,,
\end{split}
\eqlabel{d4}
\end{equation}
where in the last line we used asymptotics \eqref{h3}.
Thus 
\begin{equation}
T_3^{-1/2} \p=\frac 12 e^\r\left(1+\calo(e^{-2\r})\right)\,,
\eqlabel{d5}
\end{equation}
which upon integration of \eqref{d2} yields to leading order 
in $P^2$ and $T_3/\phi^2$
\begin{equation}
\calv_{inf}(\phi)=\phi^2\left(1-\frac{P^2}{12}\ln\frac{\phi^2}{T_3}\right)\,, 
\eqlabel{d6}
\end{equation}
which leads to the slow-roll parameter reported in \eqref{ceta}, case $(a)$.

\subsubsection{Case $(b)$}
In this case we need to find the $\calo(H^2)$ solution to
\eqref{kteqs} around KT solution \cite{kt}. Here, it is 
convenient to use $u$ as a radial coordinate. 
Analogously to the previous case we search for a solution in the 
following parametrization
\begin{equation}
\begin{split}
&K=K_{KT}+H^2 K_1\,,\quad \Phi=\Phi_{KT}+H^2 \Phi_1\,,\quad 
w=w_{KT}+H^2 w_1\,,\\
&y=y_{KT}+H^2 y_1\,,\quad z=z_{KT}+H^2 z_1\,,
\end{split}
\eqlabel{par2}
\end{equation}
where the subscript $ _{KT}$ denotes the KT solution (see eqs.~(3.9),
(3.10)
of ref.~\cite{bt})
\begin{equation}
\begin{split}
&w_{KT}=\P_{KT}=0\,,\quad e^{-4y_{KT}}=4u\,,\quad
K_{KT}=4-\frac{P^2}{2}\ln u\,,\\
&e^{-4z_{KT}}=\left(4+\frac{P^2}{2}\right)u-\frac{P^2}{2}u\ln u\,.
\end{split}
\eqlabel{ktsol}
\end{equation}
Given the parametrization \eqref{par2} we deduce from \eqref{derpot}
\begin{equation}
T_3^{-1} \frac{d\calv(u)}{du}=4 H^2 e^{4z_{KT}}
\biggl(z_1'+e^{4z_{KT}}\left(z_1 K_{KT}+\frac 14 K_1\right)\biggr)
+\calo(H^4)\,.
\eqlabel{q1}
\end{equation}
As in the example studied in \cite{bt}, it is possible to 
expand the system of equations \eqref{kteqs} around the KT 
solution \eqref{ktsol} to $\calo(H^2)$ order and obtain 
a coupled system for the deformations $\{K_1,\P_1,w_1,y_1,z_1\}$.
The resulting equations are rather complicated, and we will 
not present them here. Rather miraculously, it turns out that 
the differential equation for 
\begin{equation}
h(u)\equiv z_1'+e^{4z_{KT}}\left(z_1 K_{KT}+\frac 14 K_1\right)
\eqlabel{feq}
\end{equation}
decouples from the other equations 
\begin{equation}
0=h'-e^{4z_{KT}}K_{KT}\ h -3 e^{10y_{KT}-4z_{KT}}\,.
\eqlabel{heq}
\end{equation}
Moreover, given \eqref{ktsol}, it can be solved exactly 
\begin{equation}
h(u)=u\left(P^2(\ln u-1)-8\right)\left(c+\frac{1}{32u^{3/2}}\right)\,,
\eqlabel{hsol}
\end{equation}
where $c$ is an arbitrary integration constant. 
From \eqref{q1} we find then
\begin{equation}
T_3^{-1} \frac{d\calv(u)}{du}=-H^2\left(\frac{1}{4u^{3/2}}+8c\right) 
+\calo(H^4)\,.
\eqlabel{q2}
\end{equation}
From \eqref{infkt} we find 
\begin{equation}
\phi=\sqrt{\frac{T_3}{2}}\ u^{-1/4}+\calo(H^2)\,.
\eqlabel{fff}
\end{equation}
Finally, to leading order in $H^2$ and $T_3/\phi^2$ we find from \eqref{q2},
\eqref{fff}
\begin{equation}
\calv_{inf}(\p)=H^2\phi^2\,,
\eqlabel{last}
\end{equation}
which leads to the slow-roll parameter reported in \eqref{ceta}, case
$(b)$.


\section*{Acknowledgments}
We would like  to thank Shamit Kachru, Joe Polchinski and Johannes
Walcher for interesting discussions.
AB would like to thank KITP for hospitality 
while this work was done.  
This
research was supported in part by the National Science Foundation under
Grant No.~PHY99-07949 (AB) and PHY00-98395 (RR), as well as by the
Department of Energy under Grant No.~DE-FG02-91ER40618 (RR).



\begin{thebibliography}{99}

\bibitem{k2} 
S.~Kachru, R.~Kallosh, A.~Linde, J.~Maldacena, L.~McAllister and
S.~P.~Trivedi, ``Towards inflation in string theory,''
JCAP {\bf 0310}, 013 (2003)
[arXiv:hep-th/0308055].

\bibitem{kklt}
S.~Kachru, R.~Kallosh, A.~Linde and S.~P.~Trivedi,
``De Sitter vacua in string theory,''
Phys.\ Rev.\ D {\bf 68}, 046005 (2003)
[arXiv:hep-th/0301240].


\bibitem{ns1}
E.~Cremmer, S.~Ferrara, C.~Kounnas and D.~V.~Nanopoulos,
``Naturally Vanishing Cosmological Constant In N=1 Supergravity,''
Phys.\ Lett.\ B {\bf 133}, 61 (1983). 
J.~R.~Ellis, A.~B.~Lahanas, D.~V.~Nanopoulos and K.~Tamvakis,
``No - Scale Supersymmetric Standard Model,''
Phys.\ Lett.\ B {\bf 134}, 429 (1984).


\bibitem{ns2}
E.~Witten,
``Dimensional Reduction Of Superstring Models,''
Phys.\ Lett.\ B {\bf 155}, 151 (1985).

\bibitem{ns3}
M.~Dine, R.~Rohm, N.~Seiberg and E.~Witten,
``Gluino Condensation In Superstring Models,''
Phys.\ Lett.\ B {\bf 156}, 55 (1985).




\bibitem{gkp}
S.~B.~Giddings, S.~Kachru and J.~Polchinski,
``Hierarchies from fluxes in string compactifications,''
Phys.\ Rev.\ D {\bf 66}, 106006 (2002)
[arXiv:hep-th/0105097].

\bibitem{ks}
I.~R.~Klebanov and M.~J.~Strassler,
``Supergravity and a confining gauge theory: Duality cascades and  chiSB-resolution of naked singularities,''
JHEP {\bf 0008}, 052 (2000)
[arXiv:hep-th/0007191].


\bibitem{w96}
E.~Witten,
``Non-Perturbative Superpotentials In String Theory,''
Nucl.\ Phys.\ B {\bf 474}, 343 (1996)
[arXiv:hep-th/9604030].

\bibitem{ac}
K.~Becker, M.~Becker, M.~Haack and J.~Louis,
``Supersymmetry breaking and alpha'-corrections to flux induced  potentials,''
JHEP {\bf 0206}, 060 (2002)
[arXiv:hep-th/0204254].

\bibitem{kt}
I.~R.~Klebanov and A.~A.~Tseytlin,
``Gravity duals of supersymmetric SU(N) x SU(N+M) gauge theories,''
Nucl.\ Phys.\ B {\bf 578}, 123 (2000)
[arXiv:hep-th/0002159].

\bibitem{bt}
A.~Buchel and A.~A.~Tseytlin,
``Curved space resolution of singularity of fractional D3-branes on  conifold,''
Phys.\ Rev.\ D {\bf 65}, 085019 (2002)
[arXiv:hep-th/0111017].

\bibitem{b1}
A.~Buchel,
``Gauge / gravity correspondence in accelerating universe,''
Phys.\ Rev.\ D {\bf 65}, 125015 (2002)
[arXiv:hep-th/0203041].

\bibitem{bu} A.~Buchel, unpublished.

\bibitem{kw}
I.~R.~Klebanov and E.~Witten,
``Superconformal field theory on threebranes at a Calabi-Yau  singularity,''
Nucl.\ Phys.\ B {\bf 536}, 199 (1998)
[arXiv:hep-th/9807080].



\bibitem{blw}
A.~Buchel, P.~Langfelder and J.~Walcher,
``On time-dependent backgrounds in supergravity and string theory,''
Phys.\ Rev.\ D {\bf 67}, 024011 (2003)
[arXiv:hep-th/0207214].



\bibitem{b2}
A.~Buchel,
``Compactifications of the N = 2* flow,''
Phys.\ Lett.\ B {\bf 570}, 89 (2003)
[arXiv:hep-th/0302107].

\bibitem{td}
G.~R.~Dvali and S.~H.~H.~Tye,
``Brane inflation,''
Phys.\ Lett.\ B {\bf 450}, 72 (1999)
[arXiv:hep-ph/9812483].

\bibitem{dg}
O.~DeWolfe and S.~B.~Giddings,
``Scales and hierarchies in warped compactifications and brane worlds,''
Phys.\ Rev.\ D {\bf 67}, 066008 (2003)
[arXiv:hep-th/0208123].



\bibitem{b3}
``On effective action of string theory flux  compactifications,''
to appear. 

\bibitem{bh}
A.~Buchel,
``Finite temperature resolution of the Klebanov-Tseytlin singularity,''
Nucl.\ Phys.\ B {\bf 600}, 219 (2001)
[arXiv:hep-th/0011146].



\end{thebibliography}
\end{document}